\documentclass[11pt,aps,prd,preprint,superscriptaddress]{revtex4-1}
\usepackage{amsmath}
\usepackage[dvips]{graphicx}
\usepackage[english]{babel}
\usepackage{wasysym}

\begin{document}

\title{Phantom behavior via cosmological creation of particles}
\author{Rafael C. Nunes\footnote{E-mail:
rafaelda.costa@e-campus.uab.cat}}
\affiliation{Departamento de F\'{\i}sica, Facultad de Ciencias,
Universidad Aut\'{o}noma de Barcelona, 08193 Bellaterra
(Barcelona), Spain} \affiliation{CAPES Foundation, Ministry of
Education of Brazil, Bras\'ilia - DF 70040-020, Brazil}
\author{Diego Pav\'{o}n\footnote{E-mail: diego.pavon@uab.es}}
\affiliation{Departamento de F\'{\i}sica, Facultad de Ciencias,
Universidad Aut\'{o}noma de Barcelona, 08193 Bellaterra
(Barcelona), Spain}

\begin{abstract}
\noindent Recent determinations of the equation of state of dark
energy hint that this may well be of the phantom type, i.e.,
$w_{de} < -1$. If confirmed by future experiments, this would
strongly point to the existence of fields that violate the
dominant energy condition, which are known to present serious
theoretical difficulties. This paper presents an alternative to
this possibility, namely, that the measured equation of state,
$w_{de}$, is in reality an effective one, the equation of state of
the quantum vacuum, $w_{\Lambda} = -1$, plus the negative equation
of state, $w_{c}$, associated to the production of particles by
the gravitational field acting on the vacuum. To illustrate this,
three phenomenological models are proposed and constrained with
recent observational data.
\end{abstract}

\maketitle

\section{Introduction}
\label{sec:intro}
\noindent Recently, the analysis of  a well of data provided by
the Planck satellite \cite{ade} strengthened still further our
confidence in the, so-called, Lambda cold dark matter
($\Lambda$CDM) model. Thus far, it constitutes the most promising
cosmological model in the market because, notwithstanding its
simplicity,  it fits rather well most observational data. By
assuming  a spatially flat, homogeneous, and isotropic universe,
of which the  main sources of gravity at present are pressureless
matter (baryonic plus dark) and the energy of the quantum vacuum
(the vacuum pressure is related to the latter by $p_{\Lambda} = -
\rho_{\Lambda}$), it successfully describes, with just six free
parameters, the evolution of our Universe up to the present era of
accelerated expansion.
\\  \

\noindent A crucial quantity in models based on Einstein gravity
aimed at accounting for the current accelerated phase of expansion
is the equation of state (EoS) of dark energy (DE), the ratio
$w_{de} = p_{de}/\rho_{de}$ of the pressure to the density of dark
energy. The latter drives the acceleration thanks  to its highly
negative pressure, $p_{de} < -\rho_{de}/3$. In the case of the
$\Lambda$CDM model, this agent is nothing but the energy of the
vacuum whence the corresponding EoS parameter is just $w_{\Lambda}
= -1$. In spite of the success of this model, recent
model-independent measurements of $w_{de}$ seem to favor a
slightly more negative EoS (see, e.g., Refs. \cite{Rest},
\cite{Xia}, \cite{Cheng}, and \cite{Shafer}), which, if confirmed,
would invalidate the model. In particular, the Planck mission
yields $w_{de} = -1.13^{+0.13}_{-0.10}$ \cite{ade}.
\\  \

\noindent Rest {\it et al.},  using supernovae type Ia (SN Ia)
data from the Pan-STARRS1 Medium Deep Survey in the redshift
interval $0.03 < z < 0.65$ found $ w_{de} =
-1.166_{-0.069}^{+0.072}$, i.e., $w_{de} < -1$ at $2.3 \sigma$
confidence level \cite{Rest}. However, they caution that it is
unclear whether the tension with the $\Lambda$CDM value arises
from new physics or a conjunction of chance and hidden
systematics. Similarly, Cheng {\it et al.} obtained  $\, w_{de} =
-1.16 \pm 0.06$ ($w_{de} < -1$ at $2.6\sigma$) \cite{Cheng}.
Likewise, the authors of Ref. \cite{Shafer}, using geometrical
data from SN Ia, baryon acoustic oscillations (BAOs), and the
cosmic microwave background  (CMB) radiation, determined $w_{de} <
-1 $ at $\sim 1.9 \sigma$ level. They conclude that, at $2
\sigma$, either the Supernova Legacy Survey \cite{Conley} and
Panoramic Telescope and Rapid Response System data \cite{Scolnic}
contain unknown systematics or the Hubble constant is lower than
$71$ km/s/Mpc, or else, $w_{de} < w_{\Lambda}$.
\\  \

\noindent The simplest way to achieve $w_{de} < -1$ in a manner
consistent with general relativity  is to assume that the dark
energy corresponds to some or other scalar field, called a
``phantom field," that violates the dominant energy condition by
allowing its kinetic term to have the ``wrong" sign
\cite{Caldwell}. Thus, its energy density and pressure are given
by $\rho_{ph}= -\textstyle{1\over{2}}\, \dot{\phi}^2 + V (\phi)$
and $p_{ph} = - \textstyle{1\over{2}}\,\dot{\phi}^2 - V^2 (\phi)$,
respectively; thereby, its EoS obeys $\, w_{ph} = \rho_{ph}/p_{ph}
< -1$.
\\   \

\noindent While this kind of fields shows compatibility with
observational data -see, e.g., Ref. \cite{ade}- and constitutes a
serious contender of the $\Lambda$CDM model, its blatant violation
of the said energy condition entails serious problems on the
theoretical side. Since their energy density is not bounded from
below these fields suffer from instabilities at the classical and
quantum levels \cite{Carroll,Cline} and other maladies
\cite{Hsu,Sbisa,Dabrowski} that cast doubts about their very
existence. Nevertheless, observationally, they cannot be
discarded. Then, it is natural to wonder whether an effective EoS
less than $-1$ can be accomplished without resorting to phantom
fields. The target of this paper is to explore whether the
combined effect of the pressure of the quantum vacuum,
$p_{\Lambda} = - \rho_{\Lambda}$, and the negative creation
pressure of particles from the gravitational field of the
expanding Universe acting on the vacuum can achieve this. In this
new scenario, the effective EoS comes to be $w_{eff} = w_{\Lambda}
+ w_{c} < -1$, where $w_{c}$ is the EoS related to the creation
pressure. If the answer is in the affirmative, then the need to
recourse to phantom fields will weaken.
\\  \

\noindent This paper is organized as follows. The next section
briefly sums up the phenomenological basis of particle creation in
expanding homogeneous and isotropic, spatially flat, universes.
Section \ref{sec:scenario} proposes three different
phenomenological expressions for the particle production rate.
Section \ref{sec:data} specifies the various sets of data and the
statistical analysis employed to constrain the models. The
corresponding results are presented in Sec. \ref{sec:results}.
Section \ref{sec:comparison} briefly explores whether coupled
models of DE may give rise to effective EoS of phantom type. The
concluding section summarizes and gives comments on our findings.
As usual, a subindex zero attached to any quantity means that it
must be evaluated at present time.

\section{Cosmological models with particle creation}\label{sec:models}
\noindent As investigated  by Parker and collaborators
\cite{Parker}, the material content of the Universe may have had
its origin in the continuous creation of radiation and matter from
the gravitational field of the expanding cosmos acting on the
quantum vacuum, regardless of the relativistic theory of gravity
assumed. In this picture, the produced particles draw their mass,
momentum and energy from the time-evolving gravitational
background which acts as a ``pump" converting curvature into
particles.
\\  \

\noindent Prigogine  \cite{Prigogine} studied how to insert the
creation of matter consistently in Einstein's field equations.
This was achieved by introducing in the usual balance equation for
the number density of particles, $ (n\, u^{\alpha})_{; \alpha}=0$,
a source term on the right-hand side to account for production of
particles, namely,
\begin{equation}
(n\, u^{\alpha})_{; \alpha} = n \Gamma \, ,
\label{Eq:nbalance}
\end{equation}
where $u^{\alpha}$ is the matter fluid 4-velocity normalized so
that $ u^{\alpha}\, u_{\alpha} = 1$ and $\Gamma$ denotes the
particle production rate. The latter quantity essentially vanishes
in the radiation-dominated era (not to be considered in this
paper) since, according to Parker's theorem, the production of
particles is strongly suppressed in that era \cite{Parker-Toms}.
The above equation, when combined with the second law of
thermodynamics naturally leads to the appearance of a negative
pressure directly associated to the rate  $\Gamma$, the creation
pressure $p_{c}$, which adds to the other pressures (i.e., of
radiation, baryons, dark matter, and vacuum pressure) in the total
stress-energy tensor. These results were subsequently discussed
and generalized in Refs. \cite{Lima1}, \cite{mnras-winfried}, and
\cite{Zimdahl} by means of a covariant formalism and  further
confirmed using relativistic kinetic theory
\cite{cqg-triginer,Lima-baranov}.
\\  \

\noindent Since the entropy flux vector of matter, $n \sigma
u^{\alpha}$, where $\sigma$ denotes the entropy per particle, must
fulfill the second law of thermodynamics $(n \sigma u^{\alpha})_{;
\alpha} \geq 0$, the constraint $\Gamma \geq 0$ readily follows.
\\  \

\noindent For a homogeneous and isotropic universe, with scale
factor $a$, in which there is an adiabatic process of particle
production from the quantum vacuum, it is easily found that
\cite{Lima1,Zimdahl}
\begin{equation}
\label{presao_de_criacao} p_{c} = - \frac{\rho \, + \, p}{3H}\,
\Gamma \, .
\end{equation}
Therefore, being $p_{c}$ negative it can help drive the era of
accelerated cosmic expansion we are witnessing today. Here $\rho$
and $p$ denote the energy density and pressure, respectively,  of
the corresponding fluid; $H=\dot{a}/a$  is the Hubble factor; and,
as usual, an overdot denotes differentiation with respect to
cosmic time. Since the production of ordinary particles is much
limited by the tight constraints imposed by local gravity
measurements \cite{plb_ellis, peebles2003, hagiwara2002}, and
radiation has a negligible impact  on the recent cosmic dynamics,
for the sake of  simplicity, we will assume  that the produced
particles are just dark matter particles.

\section{Cosmic scenario}\label{sec:scenario}
\noindent Let us consider a spatially flat
Friedmann-Robertson-Walker  universe dominated by pressureless
matter, baryonic plus dark matter (DM), and the energy of the
quantum vacuum (the latter with EoS $p_{\Lambda} = -
\rho_{\Lambda}$) in which a process of DM creation from the
gravitational field, governed  by
\begin{equation}
\label{dark_matter_evolution} \dot{\rho}_{dm} \,  + \, 3H
\rho_{dm} = \rho_{dm} \, \Gamma
\end{equation}
is taking place. In writing  the last equation, we used Eq.
(\ref{Eq:nbalance}) specialized to DM particles and the fact that
$\rho_{dm} = n_{dm} \, m$, where $m$ stands for the rest mass of a
typical DM particle. Since baryons are neither created nor
destroyed, their corresponding energy density obeys
$\dot{\rho}_{b} \, + \, 3H \rho_{b} = 0$. On their part, the
energy of the vacuum does not vary with expansion, hence $\,
\rho_{\Lambda} = {\rm constant}$.
\\  \

\noindent In this scenario, the total pressure is $\, p_{\Lambda}
\, + \, p_{c}$, and thereby the effective EoS is just the sum of
the EoS of vacuum plus that due to the creation pressure,
\begin{equation}
w_{eff} = \frac{p_{\Lambda}}{\rho_{\Lambda}} \, + \,
\frac{p_{c}}{\rho_{dm}} = -1 \, - \frac{\Gamma}{3H} \, .
\label{Eq:weff}
\end{equation}
Since, by the second law, $\Gamma$ is positive semidefinite, we
have that the effective EoS can be less than $-1$ without the need
of invoking scalar fields with the wrong sign for the kinetic
term. Therefore, thanks to the combined effect of the vacuum and
creation pressures one can hope to get rid of phantom fields and
of the severe drawbacks inherent to them.
\\  \

\noindent The Friedmann equation for this scenario is
\begin{equation}
\label{fields_Einstein1}
 H^2=\frac{8 \pi G}{3} (\rho_{b} \, + \, \rho_{dm} \, + \,
 \rho_{\Lambda}).
\end{equation}
To go ahead, an expression for the rate $\Gamma$ is needed.
However, the latter cannot be ascertained before the nature of
dark matter particles be discovered. Thus, in the meantime, we
must make ourselves content with phenomenological expressions of
$\Gamma$. Here, on grounds of simplicity, we assume three
phenomenological Ans\"{a}tze, namely,
\begin{equation}
\label{proposta1} \qquad \qquad \Gamma = 3 \beta H  \qquad \qquad
\qquad \; \; \; {\rm (Model \, I)},
\end{equation}
\begin{equation}
\label{proposta2} \Gamma = 3 \beta \, H \,  [5-5 \tanh (10-12a)]
\qquad {\rm (Model \, II)},
\end{equation}
and
\begin{equation}
\label{proposta3} \Gamma = 3 \beta \, H \, [5-5 \tanh (12a-10)]
\qquad {\rm (Model \, III) } ,
\end{equation}
where $\beta$ is a constant parameter satisfying $\, 0 \leq
\beta$. As indicated, models I, II, and III correspond to
Ans\"{a}tze (\ref{proposta1}), (\ref{proposta2}), and
(\ref{proposta3}), respectively.
\\  \

\begin{figure}
\includegraphics[width=4in, height=3in]{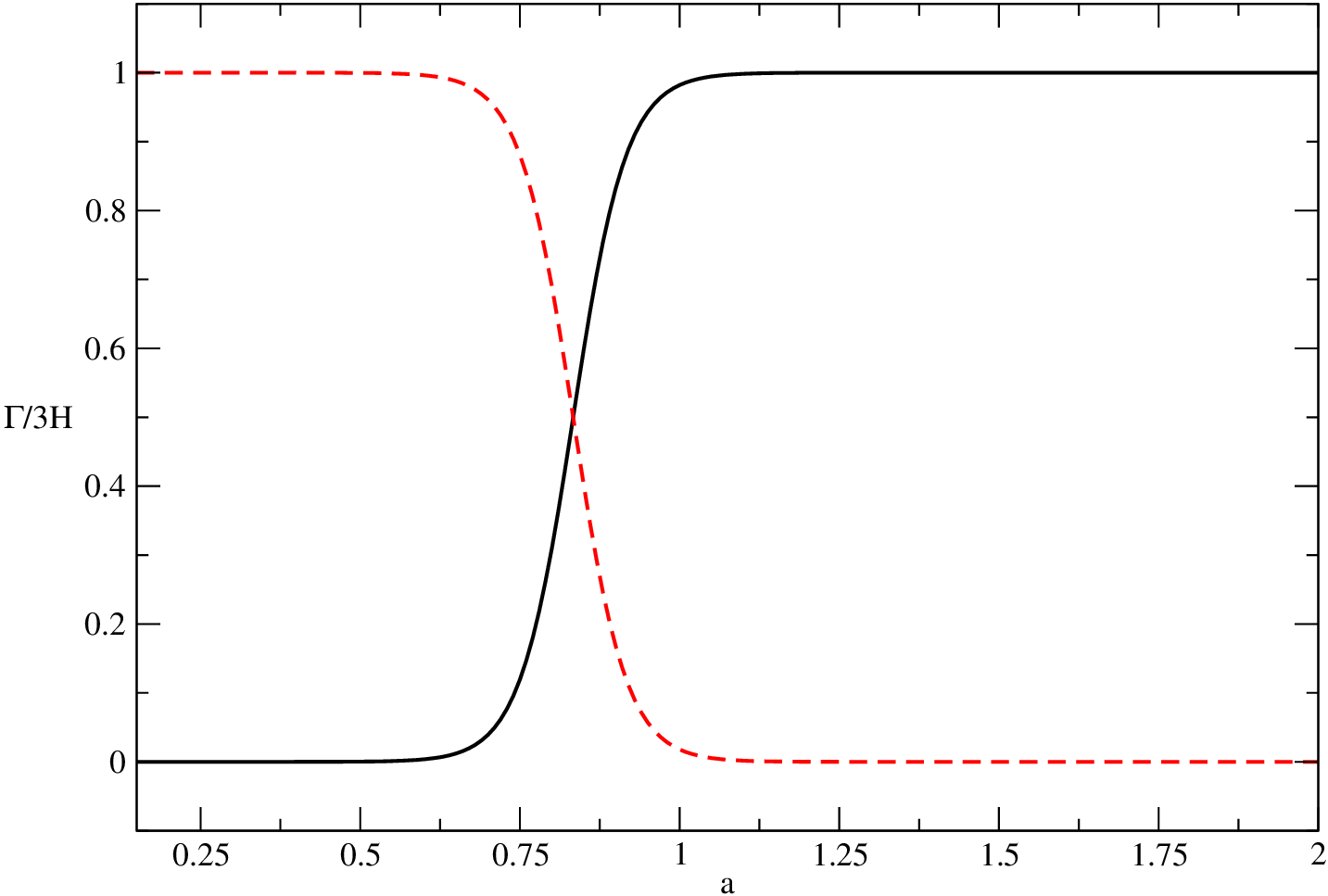}
\caption{\label{fig:gamma} Evolution of ratio $\Gamma/3H$ as a
function of the scale factor for the model II (solid, black line)
and model III (dashed, red line). In drawing the graphs we have
taken $\beta  = 0.1$.}
\end{figure}

\noindent The ratio $\Gamma/3H$ is a constant in model I. Figure
\ref{fig:gamma} shows the said ratio in terms of the scale factor
for models II and III assuming $\beta=0.1$. It should be noted
that for $\beta$ larger than $0.1$ one is led to $w_{eff} < -2 $,
which lies much away from the reported $w_{de}$ values. In all
three cases, $\Gamma/3H \leq 1$ at any scale factor.

\noindent  Inserting Eqs. (\ref{proposta1}), (\ref{proposta2}),
(\ref{proposta3}), in (\ref{dark_matter_evolution}), and
integrating, we have
\begin{equation}
\label{dark_matter_evolution_1} \rho_{dm}=\rho_{dm0} \,
a^{-3(1-\beta)}
\end{equation}
for model I and
\begin{equation}
\label{dark_matter_evolution_2} \rho_{dm}=\rho_{dm0}\, a^{-3} \,
\exp \Big(3 \beta \int_{1}^{a} \frac{g(\tilde{a})}{\tilde{a}} d
\tilde{a} \Big)
\end{equation}
for the two other models, with $g(a)=5 \, -\, 5 \tanh (10-12a)$
and $g(a)=5 \, -\, 5 \tanh (12a-10)$ for models II and III,
respectively. In both cases, $g(a)/a \rightarrow 0$ as $a
\rightarrow \infty$, whence the integral stays finite in that
limit.
\\  \

\noindent In terms of the redshift, $z=a^{-1}-1$, the Hubble
expansion rate reads
\begin{equation}
\label{Hz_model_1}
 \frac{H^2(z)}{H^2_0}=\Omega_{b0}\, (1+z)^3 \, +\, \Omega_{dm0} \, (1+z)^{3(1-\beta)} \, +
 \, \Omega_{\Lambda0}
\end{equation}
for the model I and
\begin{equation}
\label{Hz_model_2}
 \frac{H^2(z)}{H^2_0}=\Omega_{b0} \, (1+z)^3 \, + \, \Omega_{dm0}\, (1+z)^3 \,
 \exp \Big(-3 \beta \int_{0}^{z} \frac{g(\tilde{z})}{(1+\tilde{z})} d\tilde{z} \Big)\,
 + \, \Omega_{\Lambda0},
\end{equation}
for models II and III, where the $\Omega_{i0}$ denote the current
fractional densities of baryons, dark matter, and vacuum, and we
have taken into account that the baryon energy density scales as
$\, (1+z)^{3}$.  Here and throughout the scale factor is
normalized so that $\, a_{0} = 1$.

\section{Observational Constraints}\label{sec:data}
\noindent To constrain the free parameters $\theta_{i} =
(\beta,\Omega_{dm0})$ of the models above, we  use 580 Supernova
type Ia data points in the redshift interval $0.015 \leq z \leq
1.41$ of the Union 2.1 compilation \cite{SNIa}, 9 gamma-ray bursts
 in the redshift interval $1.54 < z < 3.5$ \cite{GRB2}, 6
points of baryon acoustic oscillations in the redshift interval
$0.106 \leq z \leq 0.730$ \cite{BAO}, and 28 data points of the
Hubble rate in the redshift interval $0.09 < z < 1.75$ \cite{Hz}.
Their best fit values, with their corresponding $1\sigma$
uncertainties, are presented in SEct. \ref{sec:results}. These
follow from minimizing the likelihood function $ L \propto
\exp(-\chi_{total}^{2}/2) $ with
$\chi^2_{total}=\chi^2_{SNIa}+\chi^2_{GRB}+\chi^2_{BAO}+\chi^2_{H}$,
where each $\chi_{i}^{2}$ (specified below) quantifies the
discrepancy between theory and observation. We  not use the
popular CMB shift parameter due to its comparative heavy
dependence on the standard cold matter and $\Lambda$CDM models.

\subsection{Supernovae type Ia}
\noindent Data from SN Ia are an important tool for understanding
the recent evolution of the Universe. Here, we use the Union 2.1
compilation \cite{SNIa}, available at
http://supernova.lbl.gov/Union, that contains 580 SN Ia data in
the redshift range $ 0.015 \leq z \leq 1.41$. The distance modulus
predicted for a given supernova of redshift $z$ can be expressed
as
\begin{equation}
 \mu_{th}(z)= 5 \log_{10} \left(\frac{D_{L}}{10{\rm pc}}\right) \, + \, \mu_{0}\, ,
 \label{Eq:mu}
\end{equation}
where $D_L = (1+z) \int_{0}^{z}{dz'\, \frac{H_{0}}{H(z')}}$ is the
Hubble-free luminosity distance and $\mu_{0} = 42.384 -5
\log_{10}h$, with $\, h=H_0/(100 {\rm Km/s/Mpc})$, the reduced
Hubble constant.
\\  \

\noindent Assuming the SN Ia data follow a Gaussian distribution,
we have
\begin{equation}
\label{qui_quadrado_snia}
\chi^2_{SNIa}(\theta_i,\mu_{0})=\sum_{i=1}^{580}
\frac{[\mu^{obs}(z_i)
-\mu^{th}(z_i,\theta_i,\mu_{0})]^2}{\sigma^2(z_i)},
\end{equation}
where $\mu^{obs}$ denote the observed value, $\mu^{th}$ denote the
value predicted by the model, $\theta_{i} =(\beta,\Omega_{dm0})$,
and $\sigma_{i}$ stands for the 1$\sigma$ uncertainty associated
to the $i$th data point. To eliminate the effect of the nuisance
parameter $\mu_0$, which is independent of the data points and the
data set, we minimize the right-hand side of the last equation
with respect to $\mu_{0}$ following the procedure of Ref.
\cite{prd_nesseris}. We first expand $\chi^{2}$ in terms of
$\mu_{0}$ as
\begin{equation}
\chi^{2}(\theta_{i}) = A \, - \, \mu_{0} \, B \, + \,
\mu_{0}^{2}\, C \, ,
\label{eq:expand}
\end{equation}
where
\begin{equation}
\label{qui_quadrado_snia} A(\theta_i)=\sum_{i=1}^{580}
\frac{[\mu^{obs}(z_i) -
\mu^{th}(z_i,\theta_i,\mu_0=0)]^2}{\sigma^2(z_i)},
\end{equation}
\begin{equation}
\label{qui_quadrado_snia} B(\theta_i)=\sum_{i=1}^{580}
\frac{[\mu^{obs}(z_i) -
\mu^{th}(z_i,\theta_i,\mu_0=0)]}{\sigma^2(z_i)},
\end{equation}
and
\begin{equation}
\label{qui_quadrado_snia} C=\sum_{i=1}^{580}
\frac{1}{\sigma^2(z_i)}.
\end{equation}
Equation (\ref{eq:expand}) presents a minimum for $\mu_{0} = B/C$
at
\begin{equation}
\tilde{\chi}^{2}(\theta_{i}) = A(\theta_{i})\, - \,
\frac{B^{2}(\theta_{i})}{C}.
\label{eq:tildechi}
\end{equation}
Thus, rather than minimizing $\chi^{2}(\mu_{0}, \theta_{i})$ we
minimize $\tilde{\chi}^{2}(\theta_{i})$ that is independent of
$\mu_{0}$. Clearly, $\chi^{2}_{{\rm min}} = \tilde{\chi}^{2}_{{\rm
min}}$. Thus the Hubble constant, $H_{0}$,  does not enter the
calculation of the cosmological parameters when using SN Ia data.
\\  \

\noindent In our analysis we have not taken into account the small
correlations between the SN Ia data points. As seen in the paper
by Conley {\it et al.} \cite{Conley}, the correlations induced by
systematic errors have a very minor impact (see, e.g., Fig. 11 in
that paper). This was also found by Ruiz {\it et al.},
\cite{prd_Ruiz} (see, e.g., Fig. 11 there). We feel therefore
confident that they will not significantly alter the overall
result of our analysis (i.e., that, as argued below, models I and
II may well explain the value of less than -1  reported in Refs.
[2-5] for the EoS of dark energy.)

\subsection{Gamma-ray bursts}
\noindent Gamma-ray bursts (GRBs) are very energetic astrophysical
outbursts usually at higher redshifts than SN Ia events.
Unfortunately, very often, GRBs are not standard candles as is the
case of SN Ia. Therefore, it becomes necessary to calibrate them
if they are to be employed as useful distance indicators. The
uncertainties in the observable quantities of GRBs are much higher
than in SNs Ia, since so far there is not a good understanding of
their source mechanism. These issues favor a controversy over the
use of GRBs for cosmological applications \cite{GRB1}.
\\  \

\noindent Recently, in Ref. \cite{GRB2}, a set of 9 long gamma-ray
bursts (LGRBs) in the redshift interval $1.54 < z < 3.5$ was
calibrated through the type I fundamental plane. This is defined
by the correlation between the spectral peak energy $E_p$, the
peak luminosity $L_p$, and the luminosity time $T_L \equiv
E_{iso}/ L_p$, where $E_{iso}$ is the isotropic energy. This
calibration is one of the different proposals to  calibrate GRBs
in an model cosmological-independent way. The fact that a control
of systematic errors has been carried out to calibrate these 9
LGRBs \cite{GRB2} makes this compilation compelling.
\\  \

\noindent The $\chi^2$ function for the GRB data reads
\begin{equation}
\label{qui_quadrado_grb}
\chi^2_{GRB}(\theta_i,\mu_0)=\sum_{i=1}^{9} \frac{[\mu^{obs}(z_i)
-\mu^{th}(z_i,\theta_i,\mu_0)]^2}{\sigma^2(z_i)}.
\end{equation}

Here, we perform the same procedure described in the previous
section.
\subsection{Baryon acoustic oscillations}
\noindent These can be traced to pressure waves at the
recombination epoch generated by cosmological perturbations in the
primeval baryon-photon plasma and appear as distinct peaks in the
large-scale correlation function.
\\  \

\noindent For the BAO measurements, we use the six estimates of
the BAO parameter
 \begin{equation}
  A(z)= \sqrt{\Omega_b \, + \, \Omega_{dm}} \Big[ \frac{r^2(z)}{z^2 E(z)}\Big]^{1/3}
  \label{Eq:A(z)}
  \end{equation}
given in Table 3 of Ref. \cite{BAO}, that are in the redshift
range $ 0.106 \leq z \leq  0.730$, which is the $H_0$ independent.
In this expression, $r$ is the comoving distance.
\\   \

\noindent The $\chi^2$ function for the BAO data is
\begin{equation}
\label{qui_quadrado_bao} \chi^2_{BAO}(z,\theta_i)=\sum_{i=1}^{6}
\frac{[A^{obs}(z_i) -
A^{th}(z_i,\theta_i)]^2}{\sigma^2_{BAO}(z_i)}.
\end{equation}
\\  \

\noindent Again, we neglect the correlations in the BAO data as
their impact in the final results is expected to be rather small
(see, e.g., Fig 11  in Ref. \cite{prd_Ruiz}).

\subsection{History of the Hubble parameter}
\noindent The differential evolution of early type passive
galaxies provides direct information about the Hubble parameter,
$H (z)$. An updated compilation of 28 data points $H(z)$ lying in
the redshift interval  $0.09 < z < 1.75$ can be found in Ref.
\cite{Hz}. Here, we use these data to constrain the cosmological
free parameters $\theta_{i}$ of the three models under
consideration.
\\  \

\noindent We compute the $\chi^2_{H}$ function defined as
\begin{equation}
\label{chi_quadrado_H}
\chi^2_{H}(\theta_{i},H_{0})=\sum_{i=1}^{28}
\frac{[H^{obs}(z_i)-H^{th}(z_i,H_0,\theta_{i})]^2}{\sigma^2_{H}(z_i)},
\end{equation}
where $H_{th} (z_i , H_0 , \theta_i)$ is the model-predicted value
of the Hubble parameter at the redshift $z_{i}$. This equation can
be recast as
\begin{equation}
\label{recast} \chi^2_{H}(\theta_{i},H_{0})=H^2_0\sum_{i=1}^{28}
\frac{E^2(z_i,\theta)}{\sigma^2_i} - 2H_0\sum_{i=1}^{28} \frac{
H_{obs}(z_i) E(z_i,\theta)}{\sigma^2_i} + \frac{
H^2_{obs}(z_i)}{\sigma^2_i}.
\end{equation}
\noindent The $\chi^2_H$ function depends on the model parameters
$\theta_{i}$ as well as on the nuisance parameter $H_{0}$, the
value of which is not very well known. To marginalize over the
latter, we assume that the distribution of $H_0$ is Gaussian with
standard deviation width $\sigma_{H_0}$ and mean ${\bar{H}_0}$.
Then, we build the posterior likelihood function $L_{H}(\theta)$
that depends just on the free parameters $\theta_{i}$,
\begin{equation}
\label{lH}
 L_{H}= \int \pi_{H}(H_0) e^{-\chi^2_{H}(H_0,\theta_i)} dH_0,
\end{equation}
where
\begin{equation}
\label{piH} \pi_H(H_0)= \frac{1}{\sqrt{2\pi}
\sigma_{H_0}}e^{-(H_0-\bar{H}_0)^2/2\sigma^2_{H_0}}\, ,
\end{equation}
is a prior probability function widely used in the literature. For
${\bar{H}_0}$, we take the best-fit value provided by Riess {\it
et al.} \cite{Riess}. Finally, we minimize $\chi^2_H (\theta_i) =
-2\ln L_{H}(\theta_i)$ with respect to the free parameters
$\theta_{i}$ to obtain the best-fit parameter values.

\section{Results}\label{sec:results}
Tables \ref{tab1} and \ref{tab2} summarize the main results of the
statistical analysis carried out using the set of data,
SNIa+GRB+BAO and SNIa+GRB+BAO+$H(z)$, respectively.

\begin{figure}
   \includegraphics[width=6in, height=3in]{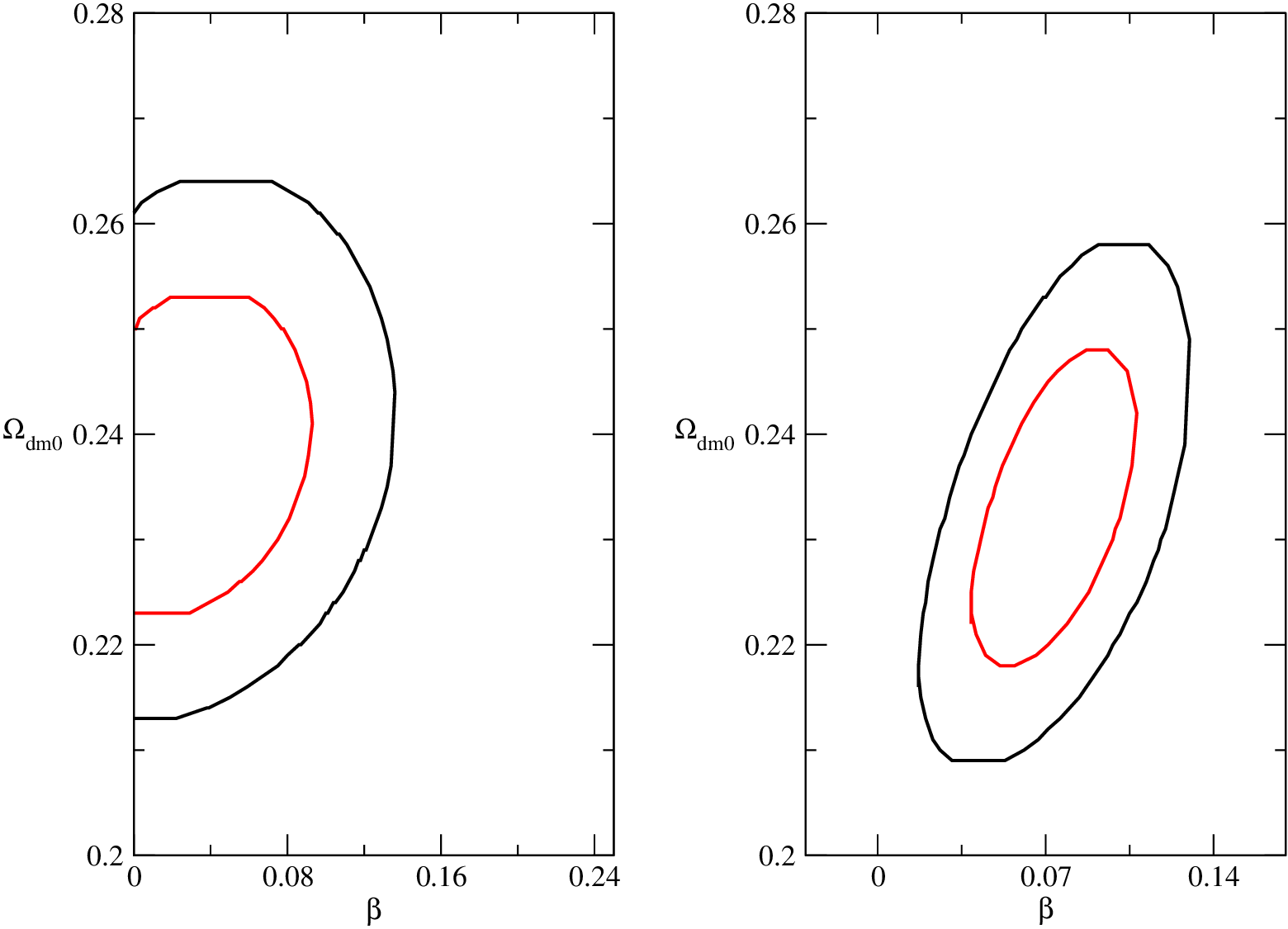}
   \caption{\label{fig:model1} Left panel: 1$\sigma$ and 2$\sigma$ confidence
  contours for model I obtained from joint analysis SNIa+GRB+BAO. Right panel:
  The same but using the set of data SNIa+GRB+BAO+$H(z)$.}
   \end{figure}

\noindent Figure \ref{fig:model1}  shows the $68\%$ and $95\%$
confidence contours in the plane $\beta-\Omega_{dm0}$ for model I
considering the sets of data SNIa+GRB+BAO (left panel), and
SNIa+GRB+BAO+$H(z)$ (right panel). Although the best fit for
$\beta$ is small for the two cases, the possibility of a small
creation rate ($\beta > 0$ $\Rightarrow$ $\Gamma > 0$) is not
ruled out. In fact, $ 0 \leq \beta \leq 0.093$ and $ 0 \leq \beta
\leq 0.136$ at 1$\sigma$ and 2$\sigma$ confidence levels,
respectively, for SNIa+GRB+BAO, and $ 0.039 \leq \beta \leq 0.108$
($ 0.017 \leq \beta \leq 0.130$) in 1$\sigma$ (2$\sigma$) when we
considered SNIa+GRB+BAO+$H(z)$. We see that a nonzero creation
rate ($\Gamma > 0 $) is consistent with the set data
SNIa+GRB+BAO+$H(z)$ in $1\sigma$ and $2\sigma$, and from this
analysis, we obtained $w_{eff}(z=0)=-1.073^{+0.034}_{-0.035}$ in
$1\sigma$ of statistical confidence. For the joint analysis
SNIa+GRB+BAO, we have an effective EoS,
$w_{eff}(z=0)=-1.0249^{+0.0249}_{-0.0691}$. This hints that the
combined effect of the quantum vacuum and the particle creation
rate may indeed result in an effective EoS lower than $-1$. The
associated uncertainty of $w_{eff}$ was determined by using the
standard error propagation method.
\\  \

 \begin{table}[!h]
     \begin{center}
         \begin{tabular}{ccccccc}
         \hline
         \hline\\
        &$\beta$&$\Omega_{dm0}$&$\chi^2_{min}/dof$\\
         \hline
         \hline\\
            Model I,   Eq. (\ref{proposta1}), &$0.025^{+0.068}_{-0.065}$&$\; \, 0.237^{+0.016}_{-0.015}$& $\; 0.9589$\\
            Model II,  Eq. (\ref{proposta2}), &$0.03^{+0.081}_{-0.069}$&$\;  \, 0.251^{+0.016}_{-0.016}$& $\; 0.9587$\\
            Model III, Eq. (\ref{proposta3}), &$0.003^{+0.007}_{-0.006}$&$\; \, 0.251^{+0.017}_{-0.016}$& $ \; 0.9590$\\
         \hline
         \hline
         \end{tabular}
     \end{center}
     \caption{Best-fit values with their $1\sigma$ errors of the free parameters,
     $\beta$ and $\Omega_{dm0}$, for the three models considered in this work
     obtained from joint analysis SNIa+GRB+BAO.}
     \label{tab1}
 \end{table}

 \begin{table}[!h]
     \begin{center}
         \begin{tabular}{ccccccc}
         \hline
         \hline\\
             &$\beta$&$\Omega_{dm0}$&$\chi^2_{min}/dof$\\
         \hline
         \hline\\
               Model I,   Eq. (\ref{proposta1}),  &$0.0729^{+0.035}_{-0.034}$&$\; \, 0.232^{+0.016}_{-0.015}$& $\; 0.9549$ \\
               Model II,  Eq. (\ref{proposta2}),  &$0.0158^{+0.0082}_{-0.0077}$&$\;  \, 0.233^{+0.015}_{-0.014}$ & $\: 0.9774$ \\
               Model III, Eq. (\ref{proposta3}),  &$0.0127^{+0.0049}_{-0.0050}$&$\;  \, 0.240^{+0.016}_{-0.015}$& $\; 0.9705$ \\
        \hline
         \hline
         \end{tabular}
     \end{center}
     \caption{Best-fit values with their $1 \sigma$ errors of the free parameters,
     $\beta$ and $\Omega_{dm0}$, for the three models considered in this work obtained
     from the joint analysis SNIa+GRB+BAO+$H(z)$.}
     \label{tab2}
 \end{table}

\begin{figure}
  \includegraphics[width=6in, height=3in]{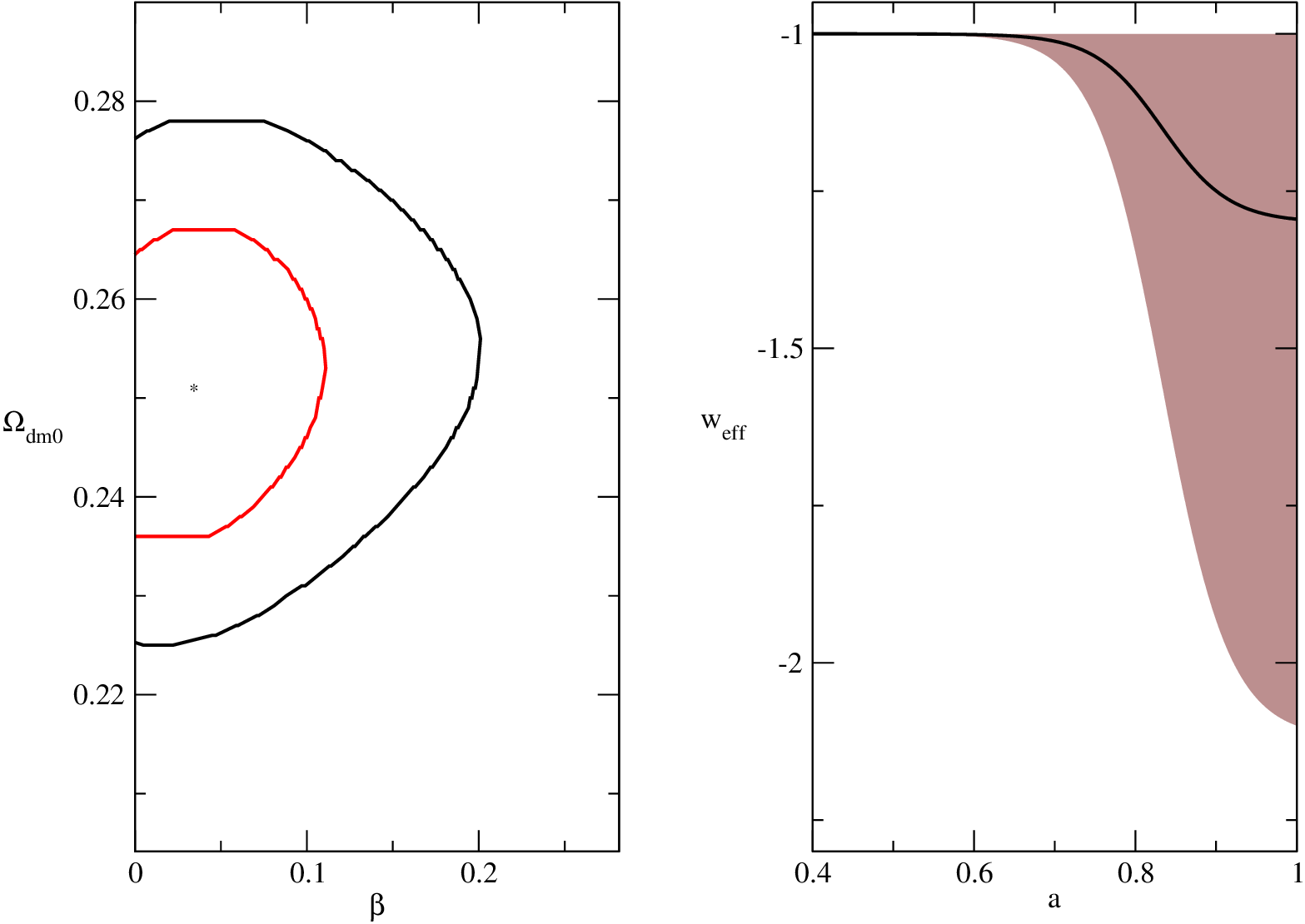}
  \caption{\label{fig:model2} Left  panel: 1$\sigma$ and 2$\sigma$ confidence
  contours for model II obtained from the joint analysis SNIa+GRB+BAO. Right panel:
  Evolution of the effective EoS in terms of the scale factor; the solid (black) line
  indicates the best-fit values, and the shaded region indicates the
  $1\sigma$ uncertainty.}
  \end{figure}

\begin{figure}
  \includegraphics[width=6in, height=3in]{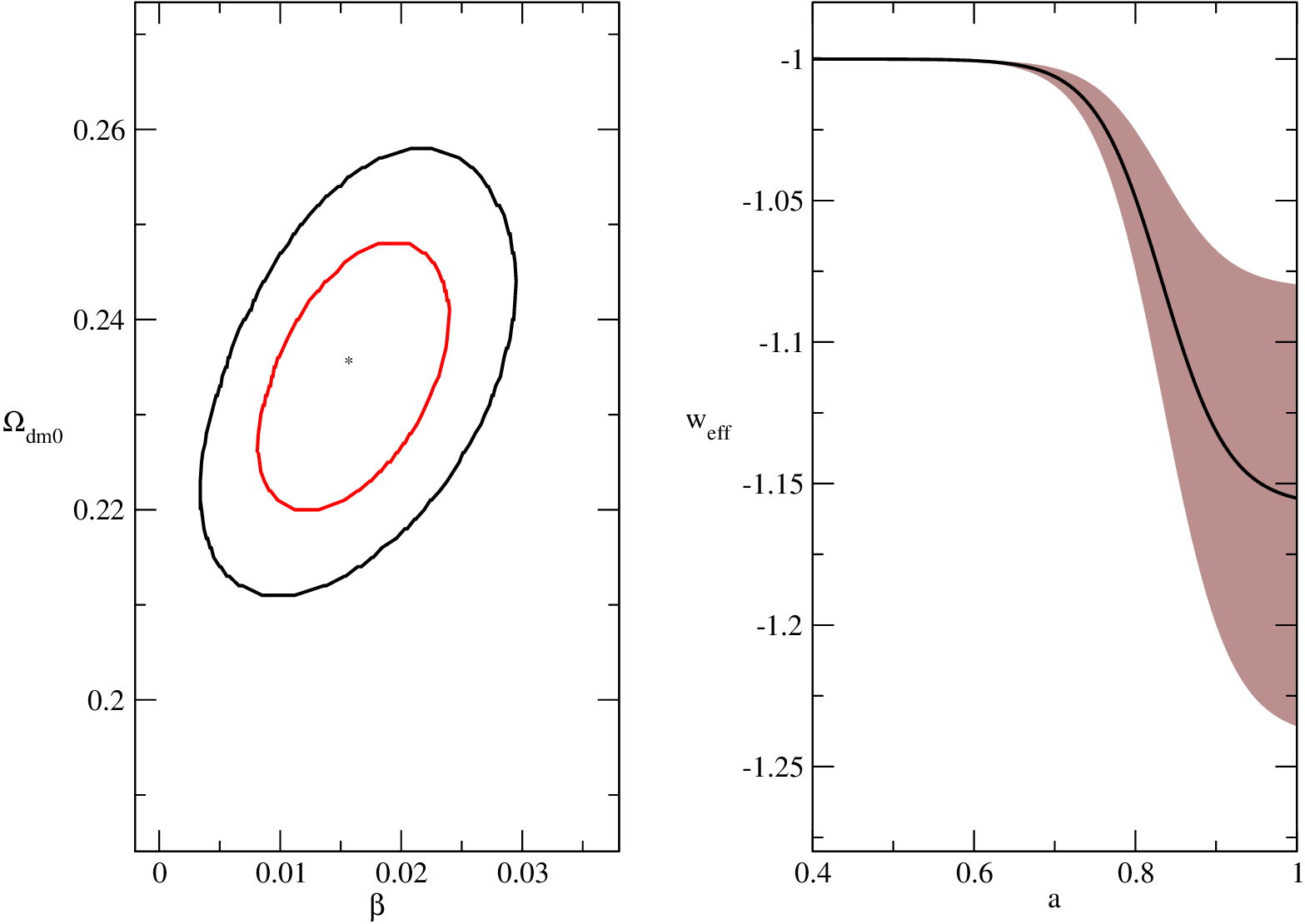}
  \caption{\label{fig:model2H} Same as Fig. \ref{fig:model2}
  but using the set of data SNIa+GRB+BAO+$H(z)$.}
  \end{figure}

\begin{figure}
 \includegraphics[width=6in, height=3in]{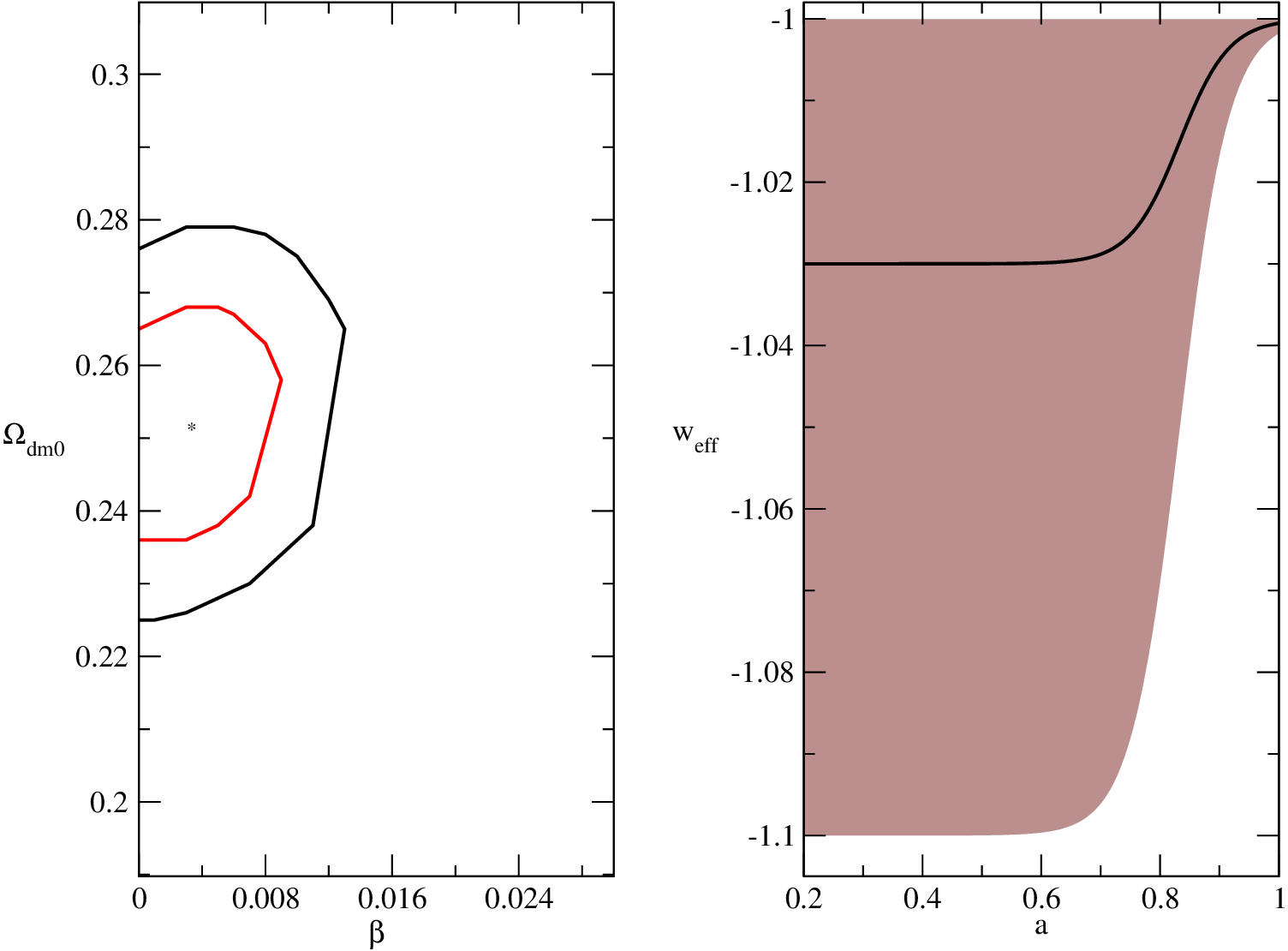}
 \caption{\label{fig:model3} Left panel: 1$\sigma$ and 2$\sigma$ confidence
  contours for model III obtained from joint analysis SNIa+GRB+BAO. Right panel:
  Evolution of the effective EoS in terms of the scale factor; the solid (black) line
  indicates the best-fit values, and the shaded region indicates the
  $1\sigma$ uncertainty.}
 \end{figure}

\begin{figure}
 \includegraphics[width=6in, height=3in]{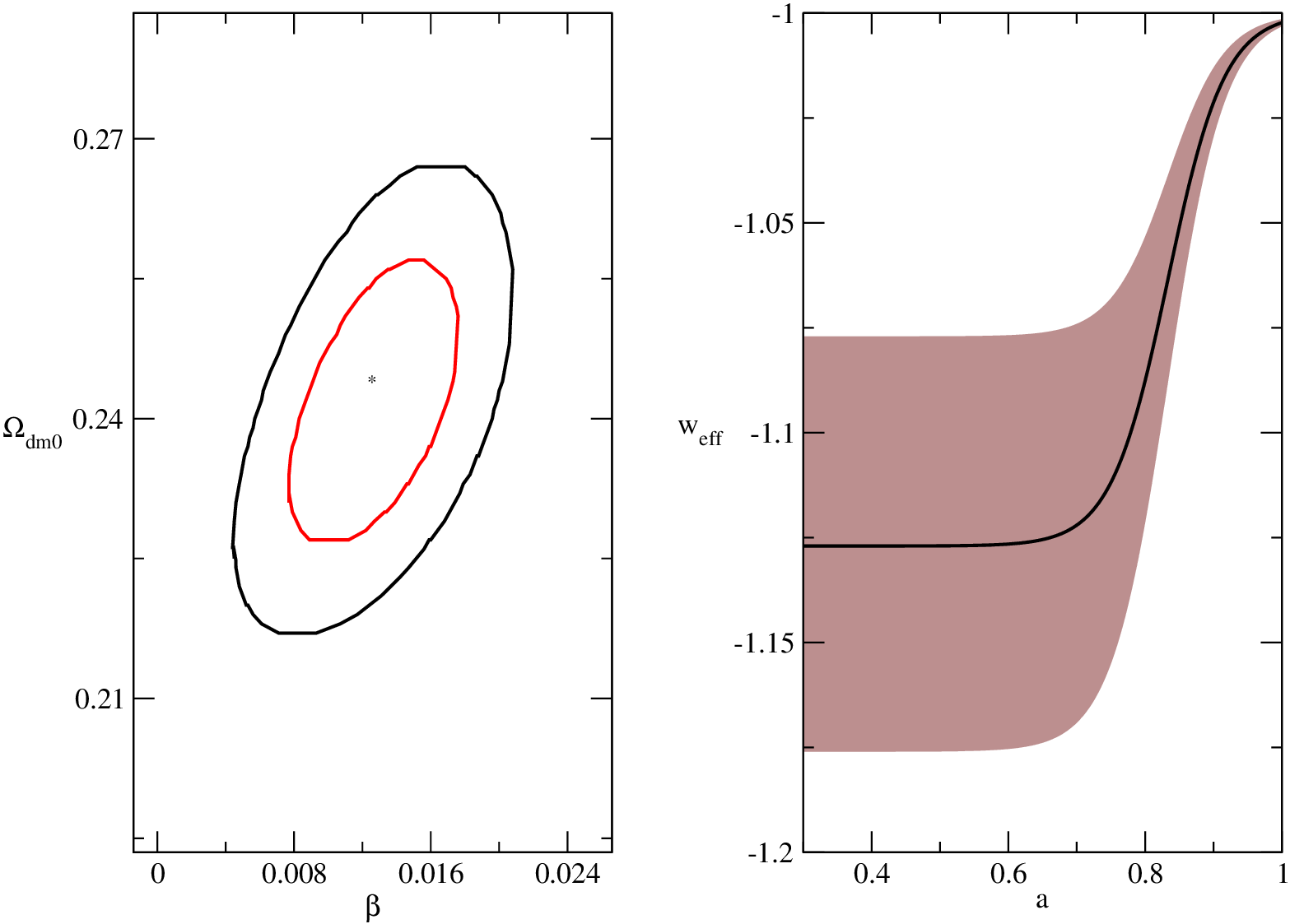}
 \caption{\label{fig:model3H} Same as Fig. \ref{fig:model3} but using the set of data
 SNIa+GRB+BAO+$H(z)$.}
 \end{figure}

\noindent Figures \ref{fig:model2} and \ref{fig:model2H} show the
$68\%$ and $95\%$ confidence contours in the $\beta-\Omega_{dm0}$
plan for models II (left panel), and the evolution of $w_{eff}$ in
1$\sigma$ (right panel) for the respective data sets,
SNIa+GRB+BAO, and SNIa+GRB+BAO+$H(z)$. As can be observed, the
inclusion of the data set, $H(z)$, significantly reduces the
constraints on the parameter $\beta$, since it, changed from $0
\leq \beta \leq 0.111$ to $0.0081 \leq \beta \leq 0.024$ in
1$\sigma$, for example. This scenario, presents as effective EoS,
$w_{eff}(z=0)=-1.2944^{+0.2944}_{-0.8049}$ for the analysis with
SNIa+GRB+BAO, and $w_{eff}(z=0)=-1.555^{+0.076}_{-0.080}$ for
SNIa+GRB+BAO+$H(z)$, both in 1$\sigma$. This model, which is
characterized by a production rate of particles that grow
throughout cosmic history, today presents a significant effective
EoS phantom. Figures \ref{fig:model3} and \ref{fig:model3H} show
the $68\%$ and $95\%$ confidence contours in the
$\beta-\Omega_{dm0}$ plan for models III (left panel) and the
evolution of $w_{eff}$ in 1$\sigma$ (right panel) for the
respective data sets, SNIa+GRB+BAO and SNIa+GRB+BAO+$H(z)$. Here,
we have that the model presents a small effective EoS in the
present moment, being $w_{eff}(z=0)=-1^{+0}_{-0.0018}$ for
SNIa+GRB+BAO and \newline $w_{eff}(z=0)=-1.002^{+0.001}_{-0.001}$
for SNIa+GRB+BAO+$H(z)$.

%

\noindent To sum up, after constraining the three models proposed
in this work with the set of observational data specified in Sec.
\ref{sec:data} it is seen first that, as a result of the joint
effect of the vacuum and the particle creation rate, effective
equations of state of phantom type can be obtained at present time
and  in the past.

\section{Comparison with models of coupled  dark
energy}\label{sec:comparison}
\noindent At this point one may wonder whether models of DE
featuring a nongravitational coupling with dark matter may also
lead to an effective EoS of phantom type. As we will see, the
answer is they very likely will not.
\\  \

\noindent The said models were proposed to lower the value of the
cosmological constant \cite{aa-wetterich} and to solve or, at
least, alleviate the coincidence problem (``why are the densities
of dark energy and dark matter of the same order precisely
today?"); see, e.g., Ref. \cite{coupled}. In any case, a coupling
between the dark components seems natural, though it must be small
as it is severely constrained by observations -see \cite{signpost}
for a review.
\\  \

\noindent For a spatially flat, homogeneous, and isotropic
universe, these models are characterized by the set of equations
\begin{eqnarray}
\dot{\rho}_{b} \,&+& \, 3H \rho_{b} = 0 \, , \\
\dot{\rho}_{dm} \,&+& \, 3H \rho_{dm} = Q \, ,\\
\dot{\rho}_{de} \,&+&\,  3H (1+w) \rho_{de} = -Q \, , \\
3 H^{2} &=& 8 \pi G (\rho_{b} \, + \, \rho_{dm} \, + \, \rho_{de})
\, , \label{eq:coupled}
\end{eqnarray}
where  $Q$ stands for the energy transfer rate per unit of volume
between the dark components and the EoS of DE is bounded by $-1
\leq w < -1/3$. This class of models implicitly assumes that the
creation pressure either vanishes or is negligible. Because of the
interaction, the energy density of DM, $\rho_{dm} = n \, m$,
differs from the usual conservation expression, $\rho_{dm} =
\rho_{dm0}\, a^{-3}$, either because $\, m \,$ varies with
expansion or because the number density of DM particles does not
obey the expression $n = n_{0} \, a^{-3}$.
\\   \

\noindent Two distinct possibilities arise: (i) $Q > 0$, which
means a continuous transfer of energy from DE to DM, and (ii) $Q <
0$ in this case, the transfer of energy would proceed in the
opposite sense. For $Q >0 $, the effective EoS, $w_{eff} = w + Q
/(3H \rho_{dm})$, results less negative than $w$, and thereby no
phantom behavior can be reproduced (this holds irrespective of
whether it is the number or mass of DM particles that varies).
\\  \

\noindent For $Q < 0 $, one follows that $w_{eff}$ can be less
than $-1$. However this kind of model can be readily told apart
from the ones discussed in the above sections. In particle
creation models, DM particles are continuously added to this
pressureless component, and thereby the amount of DM in the past
is necessarily less, at any $z > 0$, than (at the same redshift)
in the $\Lambda$CDM model that shares the same $\Omega_{b0}$ and
$\Omega_{dm0}$ than the matter creation model. Likewise, in
coupled models with $Q <0$ the amount of DM in the past was
certainly larger, at any $z > 0$, than (at the same redshift) in
the $\Lambda$CDM model with identical $\Omega_{b0}$ and
$\Omega_{dm0}$. Consequently, the growth function, $ f_{g} \equiv
{\rm d} \ln D_{+}/{\rm d} \ln a $, in matter creation models is
greater than in the $\Lambda$CDM and $f_{g}$ is also greater in
the $\Lambda$CDM than in the said kind of coupled models.
\\  \

\noindent Further, as is well known, the transfer of energy from
DM to DE worsens the coincidence problem and violates the second
law of thermodynamics \cite{grg-db}, and the conservation of
quantum numbers could be transgressed, especially if DE is the
quantum vacuum.
\\  \

\noindent In summary, models of coupled dark energy clearly
differentiate observationally from creation models,  and it is
very unlikely that they can reproduce the EoS of phantom models.

\section{Concluding remarks}\label{sec:conclusions}
\noindent Current observational data seem to favor an EoS for DE
less than $-1$ (for a quick summary of the state of art, see Fig.
1 in Ref. \cite{Shafer}). If confirmed by future experiments, like
Euclide \cite{euclide}, we will confront the puzzling situation
that  the current expansion stage of the Universe is likely driven
by a phantom field, in spite of the fact that these fields are
known to come with theoretical drawbacks of no easy fixing.
\\  \

\noindent In this paper we explored the possibility that the $\,
w_{de}$ determined by recent experiments is in reality an
effective EoS that results from adding the negative EoS, $w_{c} =
- \Gamma/3H$ (associated to the particle production pressure from
the gravitational field acting on the vacuum, \cite{Parker},
\cite{Lima1}), to the EoS of the vacuum itself, $w_{\Lambda} =
-1$. Since the rate of particle production, $\Gamma$, is not
known, we have assumed three different phenomenological
Ans\"{a}tze for the latter on grounds of simplicity, Eqs.
(\ref{proposta1})-(\ref{proposta3}), and constrained their free
parameters ($\beta$ and $\Omega_{dm0}$) of the corresponding
cosmological models with the combinations SN Ia + GRBs + BAO and
SN Ia + GRBs + BAO + H(z) of observational data sets. Here we wish
to stress that we did not fix the value of the Hubble constant,
$H_{0}$, at any point in the statistical analysis. As it turns
out, models I and II suggest that the pressure of the vacuum
combined with the particle creation pressure helps explain that
the EoS measured by recent experiments (see, e.g., Refs.
\cite{Rest}-\cite{Shafer}) is less than $-1$. Model III lies far
away from explaining it.
\\  \

\noindent To sum up, the recently reported values of lower than
$-1$ for the equation of state of dark energy may arise from the
joint effect of the quantum vacuum and the process of particle
production. This offers a viable alternative to the embarrassing
possibility of fields that, among other things, violate the
dominant energy condition, give rise to classical and quantum
instabilities, and do not respect the second law of
thermodynamics.
\\  \

\noindent Obviously, phenomenological models of particle
production different from the ones  essayed here  are also worth
exploring. More important, however, is to determine the rate
$\Gamma$ using quantum field theory but, as said above, this does
not seem feasible until the nature of DM particles is found.

\acknowledgments{ \noindent R.C.N. acknowledges financial support
from CAPES Scholarship Box 13222/13-9. This work was partially
supported by the ``Ministerio de Econom\'{\i}a y Competitividad,
Direcci\'{o}n General de Investigaci\'{o}n Cient\'{\i}fica y
T\'{e}cnica", Grant No. FIS2012-32099.}



\begin{thebibliography}{99}
\bibitem{ade} P.A.R. Ade {\it et al.} (Planck Collaboration),
``Planck 2013 results. XVI. Cosmological parameters", Astron.
Astrophys \textbf{571}, A16 (2014).
\bibitem{Rest} A. Rest {\it et al.}, ``Cosmological constraints from
measurements of type Ia supernovae discovered during the first
$1.5$ years of the Pan-STARRS1 Survey", Astrophys. J.
\textbf{795},, no 1, 44 (2014).
\bibitem{Xia} J.-Q. Xia, H. Li, and X. Zhang, Phys. Rev. D
\textbf{88}, 063501 (2013).
\bibitem{Cheng} C. Cheng, and Q.-G Huang, Phys. Rev. D
\textbf{89}, 043003 (2014).
\bibitem{Shafer} D.L. Shafer and D. Huterer, Phys. Rev. D
\textbf{89}, 063510 (2014).
\bibitem{Conley} A. Conley {\it et al.}, Astrophys. J. Suppl.
Ser.  \textbf{192}, 1 (2011).
\bibitem{Scolnic} D. Scolnic {\it et al.}, ``Systematic
uncertainties associated with the cosmological analysis of the
first Pan-STARRS1 type Ia supernova sample", Astrophys. J.
\textbf{795}, no 1, 45 (2014).

\bibitem{Caldwell} R.R. Caldwell, Phys. Lett. B \textbf{545}, 23 (2002).
\bibitem{Carroll} S.M. Carroll, M. Hoffman and M. Trodden, Phys. Rev. D \textbf{68}, 023509 (2003).
\bibitem{Cline} J.M. Cline, S. Jeon and G. D. Moore, Phys. Rev. D \textbf{70}, 86 043543 (2004).
\bibitem{Hsu} S.D.H. Hsu, A. Jenkins, and M.B. Wise, Phys. Lett. B
\textbf{597}, 270 (2004).
\bibitem{Sbisa} F. Sbisa, Eur. J. Phys. \textbf{36}, 015009
(2015).
\bibitem{Dabrowski} M. Dabrowski, arXiv:1411.2827.
\bibitem{Parker} L. Parker,  Fund. Cosm. Phys. \textbf{7}, 201 (1982);
L. Parker,  Phys. Rev. Lett., \textbf{21}, 562 (1968); L. Parker,
Phys. Rev. Lett. \textbf{183}, 1057 (1966); S.A. Fulling, L.
Parker, and B.L. Hu,  Phys. Rev. D, \textbf{10}, 3905 (1974); L.
Parker, Phys. Rev. D \textbf{17}, 933 (1978); N.J. Paspatamatiou,
and L. Parker,  Phys. Rev. D \textbf{19}, 2283 (1979).

\bibitem{Prigogine} I. Prigogine, J.  Geheniau, E. Gunzig, and P.  Nardone,
 General. Relativ. Gravit. \textbf{21}, 767 (1989).

\bibitem{Parker-Toms} L.E. Parker and D.J. Toms, {\em Quantum Field Theory in Curved Spacetime:
Quantized Fields and Gravity} (Cambridge University Press,
England, 2009).
\bibitem{Lima1} J.A.S. Lima,  M.O. Calv\~{a}o, and I. Waga,
{\em Cosmology, Thermodynamics and Matter Creation}, Frontier
Physics, Essays in Honor of Jayme Tiomno (World Scientific,
Singapore, 1990); M.O. Calv\~{a}o, J.A.S. Lima, and I. Waga, Phys.
Lett. A \textbf{162}, 223 (1992); J.A.S. Lima, A.S.M. Germano, and
L.R.W. Abramo, Phys. Rev. D \textbf{53}, 4287 (1996).

\bibitem{mnras-winfried} W. Zimdahl and D. Pav\'{o}n,
Mon. Not. R. Astron. Soc. \textbf{266}, 872 (1994).

\bibitem{Zimdahl} W.  Zimdahl, D.J. Schwarz, A.B. Balakin,
and D. Pav\'{o}n, Phys. Rev. D \textbf{64}, 063501, (2001).

\bibitem{cqg-triginer} J. Triginer, W. Zimdahl, and D. Pav\'{o}n,
Class. Quantum Grav. \textbf{13}, 403 (1996).

\bibitem{Lima-baranov} J.A.S. Lima and I. Baranov, Phys. Rev. D
\textbf{90}, 043515 (2014).

\bibitem{plb_ellis} J. Ellis, S. Kalara, K.A. Olive, C. Wetterich,
Phys. Lett. B  \textbf{228}, 264 (1989).

\bibitem{peebles2003} P.J.E. Peebles and B. Ratra, Rev. Mod. Phys.
\textbf{75}, 559 (2003).

\bibitem{hagiwara2002} K. Hagiwara {\it et al.} [Particle Data Group], Phys. Rev. D.
\textbf{66}, 010001(R) (2002).

\bibitem{SNIa} N. Suzuki {\it et al.} [The Supernova Cosmology
Project], Astrophys. J. \textbf{746}, 85 (2012); {\sf
http://supernova.lbl.gov/Union/}.

\bibitem{GRB1}  H.J. Mosquera Cuesta, H. Dumet, and  C. Furlanetto, JCAP 07(2008)004;
N. Liang, P. Wu, and S.N. Zhang, Phys. Rev. D \textbf{81}, 083518
(2010); C. Graziani, New Astron. \textbf{16}, 57 (2011); A.
Shahmoradi and  R. Nemiroff, Mon. Not. R. Astron. Soc.
\textbf{411} 1843 (2011); N.R. Butler, J.S. Bloom, D. Poznanski,
Astrophys. J. \textbf{711}, 495 (2010).

\bibitem{GRB2} R. Tsutsui {\it et al.},  arXiv:1205.2954.

\bibitem{BAO} C. Blake {\it et al.}, Mon. Not. R. Astron. Soc.
\textbf{418}, 1707 (2011).

\bibitem{Hz} K. Liao, Z. Li, J. Ming, and Z.-H. Zhu, Phys. Lett. B \textbf{718}, 1166 (2013).

\bibitem{prd_nesseris} S. Nesseris and L. Perivolaropoulos, Phys.
Rev. D \textbf{72}, 123519 (2005).

\bibitem{prd_Ruiz} E.J. Ruiz, D.L. Shafer, D. Huterer, and A Conley,
Phys. Rev. D \textbf{86}, 103004 (2012).

\bibitem{Riess} A. G. Riess  {\it et al}., Astrophys. J. \textbf{730}, 119 (2011).

\bibitem{aa-wetterich} C. Wetterich, Astron.
Astrophysics \textbf{301}, 321 (1995).

\bibitem{coupled}  L. Amendola, Phys. Rev. D
\textbf{62}, 043511 (2000); L.P. Chimento, A. Jakubi, D.
Pav\'{o}n, and W. Zimdahl, Phys. Rev. D \textbf{67}, 083513
(2003); A.A. Costa {\it et al.}, Phys. Rev. D \textbf{89}, 103531
(2014); E. Abdalla, E.G.M. Ferreira, J. Quintin, and B. Wang,
arXiv:1412.2777.

\bibitem{signpost} F. Atrio-Barandela and D. Pav\'{o}n,
``Interacting dark energy", in {\em Dark Energy -Current Advances
and Ideas}, ed. J.R. Choi (Research Signpost, Trivandrum, India,
2009).

\bibitem{grg-db} D. Pav\'{o}n and B. Wang, General Relativ. Gravit.
\textbf{41}, 1 (2009).

\bibitem{euclide} {\sf http://www.euclid-ec.org}.
\end{thebibliography}
\end{document}